\documentstyle[12pt]{article}
\begin{document}
\title{\bf PHOTOPRODUCTION OF ETAS AND N$^*$(1535) ELECTROSTRONG
PROPERTIES}
\author{NIMAI C. MUKHOPADHYAY, J. -F. ZHANG \\
{\it  Physics Department, Rensselaer Polytechnic Institute} \\
{\it Troy, NY 12180-3590, USA}\\
 M. BENMERROUCHE \\
{\it Saskatchewan Accelerator Laboratory, University of
Saskatchewan}\\{\it Saskatoon, SK S7N 0W0 Canada }}

\maketitle

\begin{abstract}
{}From the new Mainz data of eta photoproduction off
protons\cite{ref1} and deuterons\cite{ref2}, we can extract, via
the effective Lagrangian approach (ELA)\cite{ref3}, the electrostrong
 parameters $\xi_{p}$ and $\xi_{n}$ for the N$^*$(1535)
excitation. These parameters are, in units of
$10^{-4} MeV^{-1}$,  $(2.20\pm 0.15)$ and
$(-1.86\pm 0.20)$
respectively. They yield the ratio of the helicity amplitudes
for the excitation of N$^*$(1535) by photons off neutrons and
protons to be $A^{n}_{1/2}/A^{p}_{1/2}=-0.84\pm 0.15$, a
quantity of fundamental importance in the baryon structure.
Best estimates in the quark model are close to this ratio.
\end{abstract}

\section{Introduction}
There has been a revival of the study of the electromagnetic
excitation of the baryon resonances, thanks to the advent of
the CW electron machines. Current generation of these machines have
superior beam quality along with high duty factor. This,
together with good detector technology, has resulted
 in a vastly  improved data
base in the first and second resonance regions of the nucleon
spectrum. On the theoretical side, the interesting tests of QCD
involve the $Q^2$ evolution of the helicity amplitudes,
where $Q^2$ is the negative of the virtual photon mass squared.
The real photon point of this evolution curve anchors
the non-pertubative
domain, which continues till  $Q^2$ is sufficiently large to have
the perturbative QCD (PQCD) rules  set in.

The reaction of interest in this paper is the photoproduction of
the eta meson off nucleons:
\begin{equation}
\gamma +p\rightarrow\eta +p ,
\end{equation}
\begin{equation}
\gamma +n\rightarrow\eta +n ,
\end{equation}
with thresholds $E_{\gamma}=707.2 MeV$ and $706.9 MeV$
respectively, where $E_{\gamma}$ is the lab photon energy. A
large number of recent papers have looked at these reactions both
experimentally\cite{ref1,ref2} and
theoretically\cite{ref3,ref4,ref5,ref6,ref7}. A particular
 interest in this reaction is to study the nucleon to
N$^*$(1535) helicity amplitude that dominates this\cite{ref3}
class of reactions. On the experimental side,  very high
quality experiments  have been done at Mainz\cite{ref1,ref2},
while less accurate investigations at Bates\cite{ref8} and
Bonn\cite{ref9} have also added to our
knowledge in this region.
Although the theoretical investigations have been principally in
the context of quark models, sophistications have been
introduced in the framework of these models\cite{ref10} making a
meaningful comparison between theory and experiment possible.

A particular interest from the theoretical point of view is the
extraction of the amplitudes $A^{p}_{1/2}$ and $A^{n}_{1/2}$
for the p and n targets respectively to excite the
N$^*$(1535) resonance. Granted that we  have to deal with
model dependence in this extraction from the data, it is of
value to see how far this model dependencce can be reduced.

In our analysis, we construct the  photoproduction
amplitude by using the ELA\cite{ref3},
 which
has been very successful in the first\cite{ref11},
 second\cite{ref3} and third\cite{ref12}
resonance regions. The main contributions\cite{ref3}
 to the amplitude
are the nucleon Born term, vector meson exchanges and
the excitation of the  nucleon resonances,
dominated in this case by the N$^*$(1535) resonance. In
contrast to the pion photoproduction, where the pion-nucleon
channel couples to a large number of s-channel resonances,
the eta-nucleon channel has strong selectivity of the
N$^*$(1535) intermediate state.

\section{The $E_{0^{+}}$ amplitude}

The dominant amplitude in the N$^*$(1535) excitation region is
the $E_{0^{+}}$ amplitude. In Table 1, we show the real
part of the $E_{0^{+}}$ amplitude at the threshold of the
eta photoproduction, as obtained from our ELA fits to the
various sets of data. Notice the large difference between the
$E_{0^{+}}$ amplitude as obtained from our fits to the old
data base\cite{ref13} and the results inferred from our fits
to the photoproduction data obtained at Mainz, from the
proton\cite{ref1} and the neutron\cite{ref2} targets, by
Krusche {\it et al.}. The latter are not direct measurements,
but inferences from the photoproduction off deuteron
targets\cite{ref2}. Notice the sign difference between
$E^{p}_{0^{+}}$ and $E^{n}_{0^{+}}$, and the dominance of
the N$^*$(1535) in both of these amplitudes. This is the
basis for our extraction of the electrostrong properties
of the N$^*$(1535) excitation off nucleons.
\begin{table}
\caption{A comparison of various contributions to
  the $E_{0^+}$ multipole, in
units of $10^{-3}/m_{\pi+}$, for the $\gamma +p\rightarrow\eta +p$ and
the $\gamma +n\rightarrow\eta +n$
reactions, at their respective thresholds. The parameters, defined in
\protect{\cite{ref3}}, $\alpha =1$ and
 $\Lambda^{2}=1.2 GeV^{2}$ are used.
The targets  are indicated.
 Our model parameters are fitted to the
experiment of Krusche {\it et al.}\protect{\cite{ref1}} for protons,
and our inferred angular distributions
 for neutrons\protect{\cite{ref2}}. }
\label{table1}
\begin{center}
\begin{tabular}{lccc}
\hline\hline
 &\multicolumn{1}{c}
{Old data base}&\multicolumn{2}{c}{Mainz Data}\\
      & p &p & n  \\[1em] \hline

Nucleon Born terms &$-6.05$&$-3.4$ &$2.3$
\\[1em]
$\rho+\omega$      &$2.89$ & $2.9$ &$-2.0$
 \\[1em]
$N^{*}(1440)$      &$-0.47$ &$--$ &$--$ \\ [1em]
$N^{*}(1535)$      &$12.06$&$12.2$& $-10.5$
 \\[1em]
$N^{*}(1520)$      &$1.46$&$1.6$&  $-0.2$\\[1em]
$N^{*}(1650)$      &$-0.94$ &$--$ &$--$ \\ [1em]
$N^{*}(1710)$      &$0.25$ &$--$ &$--$ \\ [1em]
Total              &$9.21$&$13.3$ & $-10.4$\\[1em]
\hline \hline
\end{tabular}
\end{center}
\end{table}

\section{The differential  eta photoproduction
cross-section}

In Fig.1  we present a sampling of the angular
distributions, as obtained by Krusche {\it et al.} for the proton
target and   the angular distributions  for the neutron
target as inferred from their data off the deuteron
target. Fig.1, upper pannel,
shows our fit to the proton data for parameters $\alpha =1.0$,
$\beta=0.88$, $\delta =2.05$ for the N$^*$(1520) sector\cite{ref3}
and the parameter $\Lambda^{2}=1.2 GeV^2$ for the vector meson
sector\cite{ref3}. Fig.1, lower pannel,  incorporates
 the prediction  of the Capstick  quark
model\cite{ref10} at the same $\alpha$, $\beta$, $\delta$ and
$\Lambda$.
The quality of fit does not change appreciably when
these parameters are varied. The flatness of the angular
distribution at low energy is a direct reflection of the
dominance of the $E_{0^{+}}$ amplitude, which, in turn, is
dominated by the N$^*$(1535) excitation.
 The quality of these
comparisons is consistently excellent.

\section{The extraction of the electrostrong amplitude
parameter for the N$^*$(1535) excitation}

As Benmerrouche and Mukhopadhyay\cite{ref3} first showed,
one can reduce the model dependence of the ELA in fitting the data
by introduction of the parameter
\begin{equation}
\xi_{i}=\sqrt{\chi_{i}\Gamma_{\eta}}A^{i}_{1/2}/\Gamma_{T}
\end{equation}
where $\chi_{i}$ is a kinemetic factor, $\Gamma_{\eta}$ and
$\Gamma_{T}$ are the eta-nucleon and total widths of the
N$^*$(1535) decay, $A^{i}_{1/2}$ is the electromagnetic
helicity amplitude in the target $i$ ($i=p$ or $n$) to
excite N$^*$(1535) resonance.

In Table 2, we present three  different fits in our ELA to
the Mainz proton data to establish the following fact:
{\it while the extraction of the $A^{p}_{1/2}$ is
model-dependent and is strongly dependent on the choice of
$\Gamma_{\eta}$ and $\Gamma_{T}$, the parameter $\xi_{p}$
is not.} We can repeat this excercise\cite{ref3} and show the
same to be true for the neutron target. Thus, from the recent Mainz
data, our ELA fits yield the following parameters, which
are largely insensitive to the model uncertainties of our
ELA,  also confirmed by similar numbers obtained in
a simple Breit-Wigner fit of the N$^*$(1535) excitation of
the $E_{0^{+}}$ multipoles, inferred from the experiment\cite{ref3}.
We get
\begin{equation}
\xi_{p}=(2.20\pm 0.15)\times 10^{-4} MeV^{-1},
\end{equation}
\begin{equation}
\xi_{n}=(-1.86\pm 0.20)\times 10^{-4} MeV^{-1},
\end{equation}
yielding the ratio
\begin{equation}
A^{n}_{1/2}/A^{p}_{1/2}=-0.84\pm 0.15 .
\end{equation}
Notice that {\it all strong interaction effects have
dropped out in
extracting the last ratio.} Thus, this is a quantity
particularly  suitable for a direct comparison with the
hadron models.
\begin{table}
\caption{Fitted $A^{p}_{1/2}$
 and inferred $\xi_{p}$ for N$^*$(1535)
obtained from the new MAMI data\protect\cite{ref1}, along with
$\chi^2$ per degree of freedom, for various analyses.
Three rows are fits to the
differential cross sections,with different values of strong
eta-nucleon widths and resonance parameters.
a and b have N$^*$(1535) amplitude only while c consists
of the ELA amplitude as described in the text.}
\begin{center}
\begin{tabular}{lccc}
\hline \hline
&$A_{1/2}$&$\xi$ &$\chi^2/df$   \\[1em]
\hline
a& $113$&2.2 & 2.8   \\[1em]
b&$144$& 2.3 &3.0   \\[1em]
c&$98$& 2.2& 1.4 \\[1em]
\hline\hline
\end{tabular}
\end{center}
\end{table}

\section{Concluding remarks}

We do not have yet any ``measurement" of this ratio or the
estimates of the parameters $\xi_{i}$ in the lattice QCD
approach, but that would be very timely, and welcome.
 In the meantime, we can
compare the amplitudes $\xi_{i}$ and $A^{n}_{1/2}/A^{p}_{1/2}$
with the available quark model results\cite{ref10}. These
results are qualitatively (and some cases quantitatively)
in agreement with our extracted value, but
more precision is needed in the fomer to compare with
the precision  we can
get out of the data.

Polarization observables are going to be very helpful
in futher tests of our extracted parameters. New facilities
like the Mainz microtron(MAMI), Grenoble light source
(GRAAL) and the Continuous Electron Beam Accelarator Facility
(CEBAF) will help these developments in the near future.

We thank Prof. B. Krusche for many helpful conversations.


Fig. 1: {The  upper pannel: the angular distributions for eta
photoproduction off protons for two sample values of $E_{\gamma}$,
716 and 775 MeV.
Experimental points (circles) are from \protect{\cite{ref1}}, and the
dot-dashed lines are our  effective Lagrangian  fits.
The lower pannel:  our
predictions for the neutron target from quark model
using  $A^{n}_{1/2}/A^{p}_{
1/2}=-0.83$ (solid line), vs. the
inferred differential cross-section (stars)\protect{\cite{ref2}}.}

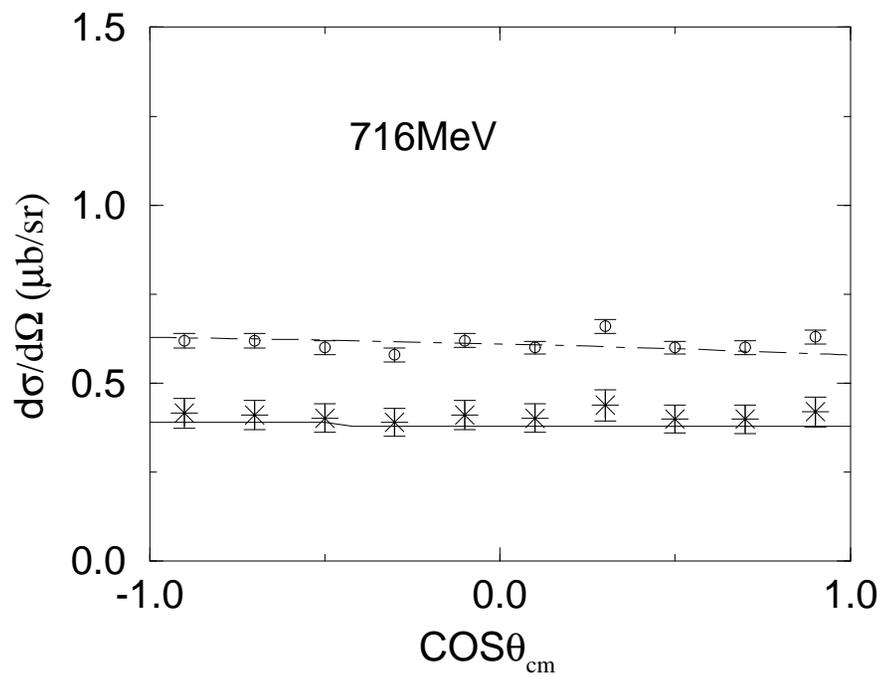
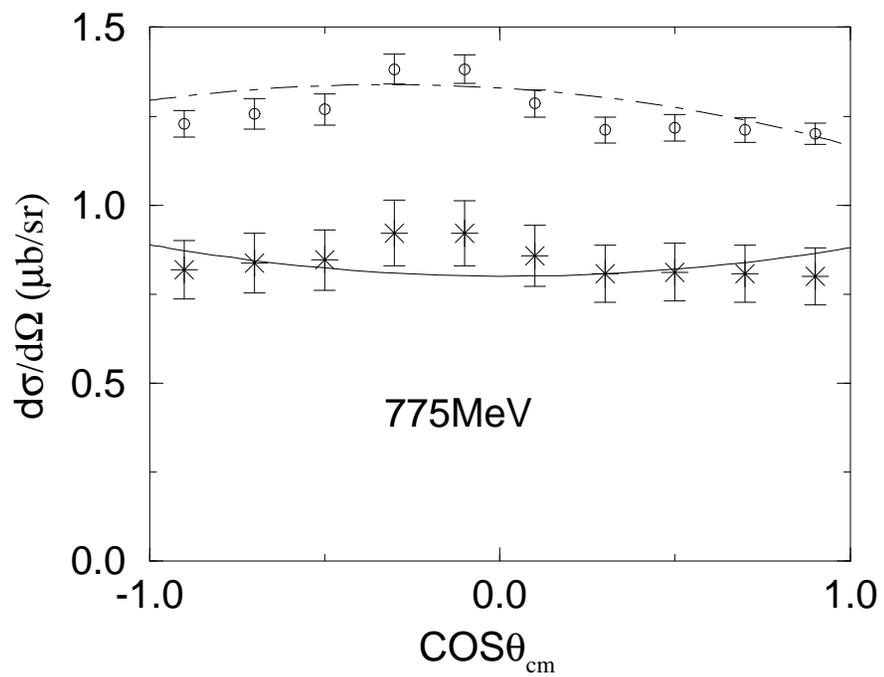
\end{document}